\documentclass[oribibl,fleqn,a4paper,12pt]{llncs}
\usepackage[latin1]{inputenc}
\usepackage[T1]{fontenc}
\usepackage[ngerman,english]{babel}
\usepackage{graphicx}
\usepackage{pifont}
\usepackage{textcomp}
\usepackage{amsmath}
\usepackage{amssymb}
\usepackage[final]{pdfpages}
\usepackage{isabelle}
\usepackage{isabellesym}
\usepackage{xspace}

\newcommand{\pisastyle}{\normalsize\tt\slshape}
\newcommand{\spisastyle}{\small\tt\slshape}
\newcommand{\alice}{\textsf{ALICE}\xspace}
\newcommand{\Focus}{\textsc{Focus}\xspace}
\newcommand{\io}{{I/O}$^*$\xspace}

\newcommand{\sisanewline}{\mbox{}\protect\\*[-1.6ex]\mbox{}}
\newcommand{\makespace}{\protect\vspace{-0.4cm}}
\newcommand{\makedoublespace}{\protect\vspace{-1cm}}
\newcommand{\delsecspace}{\protect\vspace{-0.2cm}}

\begin{document}
\pagestyle{empty}


\title{\alice\\
An Advanced Logic for Interactive Component Engineering}
\author{Borislav Gajanovic \and Bernhard Rumpe}
\institute{Software Systems Engineering Institute \\
Carl-Friedrich-Gauß Faculty for Mathematics and Computer Science \\
Braunschweig University of Technology, Braunschweig, Germany \\
http://www.sse-tubs.de}
\maketitle

\begin{abstract}
This paper presents an overview of the verification framework \alice
in its current version 0.7.
It is based on the generic theorem prover Isabelle \cite{Pau03a}.
Within \alice a software or hardware component is specified as a state-full
black-box with directed communication channels.
Components send and receive asynchronous messages via these channels.
The behavior of a component is generally described as a relation on the
observations in form of streams of messages flowing over its input and
output channels. Untimed and timed as well as state-based, recursive,
relational, equational, assumption/guarantee, and functional styles of
specification are supported.
Hence, \alice is well suited for the formalization and verification of
distributed systems modeled with this stream-processing paradigm.
\end{abstract}
\delsecspace
\delsecspace
\delsecspace

\delsecspace
\section{Introduction}\label{sec:intro}

\delsecspace
\subsection{Motivation}\label{subsec:motivation}

As software-based systems take ever more and more responsibility in this world,
correctness and validity of a software-based system is increasingly
important. As the complexity of such systems is also steadily increasing,
it becomes ever more complicated to ensure correctness.
This especially concerns the area of distributed systems like bus
systems in transportation vehicles, operating systems, telecommunication
networks or business systems on the Internet.
Expenses for verification are an order of magnitude higher than
the expenses of the software testing up to now. 
This, on the one hand, will not change easily in the short run 
but it will also become evident that crucial parts
of software need a different handling than less critical ones.
So verification will go along with testing in the future. Full
verification, however, will at least be used for critical protocols and
components. To reduce verification expenses, a lot has been achieved
in the area of theorem provers, like Isabelle
\cite{Pau03a, Pau03b, NPW02},
in the last years. Based on these foundational works and on the increasing
demand for powerful domain specific theories for
such theorem provers, we have decided to realize \alice as
a stream-processing-oriented, formal framework for distributed, asynchronously
communicating systems.

\begin{figure}
\begin{center}
\scalebox{0.5}{\includegraphics{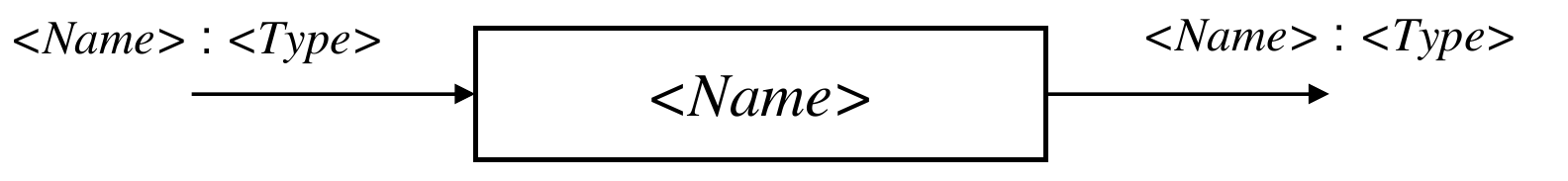}}
\caption{Illustration of an interactive component as a black-box}
\label{fig:icomponent}
\end{center}
\end{figure}

\alice is a still growing framework within Isabelle for the verification of
logically or physically distributed, interactive systems,
where the concept of communication or
message exchange plays a central role.
An interactive system (see also \cite{BS01} for a characterization) 
consists of a number of components with precisely defined interfaces.
An interactive component interacts with its environment 
via asynchronous message sending and receiving over directed and 
typed communication channels. 
Each channel incorporates an implicit, unbounded buffer that 
decouples the sending and arrival of messages, and thus describing 
asynchronous communication. 
In timed channels, we can control how long these messages
remain in this implicit buffer.
Fig. \ref{fig:icomponent} illustrates the graphical notation for the 
syntactical interface of a simple interactive component with one input and 
one output channel.

\makespace
\makespace
\vspace{-0.2cm}

In \alice message flow over channels is modeled by possibly infinite
sequences of messages called streams. Such a stream represents the 
observation of what happens on a channel over time.
Since infinite sequences are also
included, the liveness and fairness properties of systems can also be dealt
with. \alice provides type constructors {\pisastyle{astream}}
for building (untimed) streams and {\pisastyle{tastream}} for timed streams
over arbitrary types.

As an advanced verification framework, \alice will offer precisely formalized
and comfortably usable concepts based on an underlying logic language
called HOL \cite{NPW02} as available in Isabelle.
Using a well explored and rather expressive logic language
allows us to build on an elaborated set
of verification techniques for \alice.

\alice will provide support for a number of
techniques to specify a component. A specification
can be a relation between input and output streams,
a stream-processing function, a mapping of input to output,
or a set of stream-processing functions allowing to describe
non-determinism and underspecification.
All variants can be timed or untimed. Further support will be given
to map between these styles, allowing to choose appropriate
specification techniques for each problem and integrating those later.

Although \alice does already provide some of these features 
in its current version, this workshop
paper also reports on work still to be done 
(for the previous version see \cite{GR06}).
In a future version \alice will provide the following:

\begin{itemize}

\item A verification framework based on Isabelle supporting development
methods for real time, distributed, asynchronously communicating and
object oriented systems, respectively. This supports e.g.
the development methodologies of \cite{Rum96} and \Focus \cite{BS01}.

\item A formal semantics framework
for various languages based on \linebreak[4] 
stream-processing, e.g. UML's composite
structure diagrams that will be formalized based on streams
\cite{BCR06, BCR07a, BCR07b}.

\item A sophisticated verification tool for distributed, interactive systems
or, at least, their communication protocols based on stream-processing
(see \cite{Ste97} for a survey of stream-processing).

\end{itemize}

In the following we
give a compact overview of Isabelle's HOL and HOLCF that acts as a reminder
for experts of the field. An introduction can be found in
\cite{NPW02, Reg94a} before we start describing features of \alice
in Section \ref{sec:alice} and demonstrating the use of \alice in Section
\ref{sec:abp} on the Alternating Bit Protocol. Section \ref{sec:lastsec}
concludes the paper with a discussion.

\delsecspace
\subsection{HOL}\label{subsec:hol}

Isabelle is a generic theorem prover, hence,
it can be instantiated with object logics and
appropriate proof tactics.
Isabelle/HOL \cite{NPW02}, in short HOL,
is such an elaborated higher order logic, dealing amongst others with sets,
relations, total functions, natural numbers, and induction.

HOL provides a term syntax close to
mathematical syntax and constructs from functional languages.
It also provides basic types like {\pisastyle{bool}} or {\pisastyle{nat}}.
For building sets over arbitrary types, HOL provides the type constructor 
{\pisastyle{set}}.
Function types are built with the infix type constructor
$\Rightarrow$ for total functions. To build more complex
types, the mentioned, and a number of additional
basic types and type constructors are provided.

HOL inherits the type system of Isabelle's metalogic
including its automatic type inference for variables.
There are two kinds of implication:
the implication on the level of object logic, in this case HOL,
symbolized by $\longrightarrow$, and the symbol $\Longrightarrow$
for Isabelle's inference. Analogously, there is an object logics symbol
for the equality, in this case $=$, and the metalogics symbol $\equiv$ for
the definitional equality.

In Isabelle  assumptions of an inference rule are enclosed in
\textnormal{\textlbrackdbl} \textnormal{\textrbrackdbl}
and separated by \verb|;|. The metalogics 
universal quantifier is symbolized by
$\bigwedge$.

\delsecspace
\subsection{HOLCF}\label{subsec:holcf}

Isabelle/HOLCF \cite{Reg94a, MNvOS99}, shortly HOLCF,
is a conservative extension of
HOL with domain theoretical concepts, such as chains, continuity,
admissibility, fixpoint recursion and induction, as well as some basic 
types and datatypes e.g. for lazy lists.

HOLCF extends HOL with the central type class {\pisastyle{pcpo}}
for ``pointed complete partial orders''. Any type that is a member
of this type class features a special relation symbol
{\isasymsqsubseteq} for a partial order on its elements, the least element
symbolized by $\bot$, and the existence of the least upper bound for any
chain of its elements with respect to {\isasymsqsubseteq}.

This extension is carried out in layers of theories,
beginning with the definition of type class {\pisastyle{po}}
for partial orders. {\pisastyle{po}} is extended to
type class {\pisastyle{cpo}}, where the existence of the least upper bound for
any chain, symbolized by {\isasymSqunion}{\pisastyle{i. Y i}}, is introduced.
Here, {\pisastyle{Y}} is a chain of growing elements and {\pisastyle{i}}
the index over natural numbers.
Based on these theories, monotonicity and continuity for HOL functions on 
{\pisastyle{cpo}} types is formalized. 

Type class {\pisastyle{pcpo}} finally introduces the
existence of the least element in its members. 
We call the members of this class HOLCF types.
Subsequently, HOLCF provides the new infix 
type constructor $\rightarrow$ for the construction of continuous functions 
on HOLCF types. Analogously to the HOLCF types, 
we call these functions HOLCF functions or operations.
These functions, by definition, exhibit the advantages
of continuous functions, such as composability, implementability etc.
A lambda-abstraction, denoted by
$\Lambda$ (not to confuse with HOL's $\lambda$) and a corresponding function
application, using the symbol {\isasymcdot}
(opposite to HOL's white space) is provided accordingly.

Subsequently, the fixpoint theory {\it {Fix}}
mainly implements a continuous fixpoint operator, symbolized by
{\pisastyle{fix}}, and the fixpoint induction principle.
Hence, with $\rightarrow$, {\pisastyle{fix}}, and HOLCF datatypes a complete 
HOLCF syntax for defining and reasoning about 
HOLCF functions and types is provided, which is separate from
HOL's function space. As an advantage, by construction, HOLCF function
abstraction and application remains in the HOLCF world.

\delsecspace
\subsection{Related Work}

A good outline on different approaches to formalize possibly infinite
sequences in theorem provers like Isabelle or PVS, 
as well as a detailed comparison can be found in \cite{DGM97, Mül98}.
In contrast to a HOLCF formalization given in \cite{Mül98},
where finite, partial, and infinite sequences are defined to model
traces of I/O-Automata, our streams have been developed 
using only partial sequences and their infinite completions, 
which are more appropriate for modeling interactive systems as these
are generally non-terminating. A pure HOL approach based on 
coinduction and corecursion is described in \cite{Pau97}.

Another approach is 
the formal specification language ANDL introduced in \cite{SS95}.
ANDL is a formalization of a subset of \Focus with an untimed syntax and 
a fixed and an untimed semantics. 
Currently, ANDL does not provide an appropriate verification infrastructure
or extended sophisticated definition principles, but it is HOLCF oriented.
In \cite{SM97} ANDL is used as interface for an A/C refinement 
calculus for \Focus in HOLCF. 
In \cite{Hin98} ANDL is extended to deal with time.

A recent work in this area is \cite{Spi06}, where a pure HOL approach to 
formalize timed \Focus streams is used. By this approach 
(see also \cite{DGM97, Mül98}), an infinite stream is represented by a 
higher-order function from natural numbers to a set of messages.
Furthermore a time-driven approach, as it will briefly be mentioned in Section 
\ref{subsec:proofprinciples}, has been chosen there.

Apart from our idea of building such a logical framework, the realization 
of \alice is based on
a rudimentary formalization of \Focus streams in HOLCF, 
developed by D. von Oheimb, F. Regensburger, and B. Gajanovic 
(the session HOLCF/FOCUS in Isabelle's release Isabelle2005),
a concise depiction of HOLCF in \cite{MNvOS99},
as well as on the conclusions from \cite{DGM97, SM97}.
It is elaborately explained in \cite{GR06}.
Additionally, it is worth mentioning that, 
in the current version HOL's construct 
{\pisastyle{typedef}} has been used to define {\pisastyle{astream}}.

\delsecspace
\section{\alice}\label{sec:alice}


The newly defined logic \alice includes the following parts:

\begin{itemize}

\item HOL - the full HOL definitions.

\item HOL/HOLCF - all theories from HOLCF,
like {\it{Pcpo}}, {\it{Cont}}, etc. that are used on the ``interface''
between HOL and HOLCF (as discussed in Section \ref{subsec:holcf}).

\item HOLCF - using HOLCF application/abstraction (LCF sublanguage) only.

\item \alice\ - basic type constructors {\pisastyle{astream}} and
{\pisastyle{tastream}}, as well as recursion, pattern-matching, automata, etc.

\item \alice\ - lemmas provided by \alice theories (they are generally
partitioned in timed and untimed properties).

\end{itemize}

Please note that, for the development of \alice, we use a
combination of HOL and HOLCF syntax, but the user of \alice 
does not need to.
This is due to the fact that we internally use HOLCF to build up necessary
types, operators, and proving techniques, but will encapsulate these
as much as possible. 

\delsecspace
\subsection{Basic Features of \alice}\label{subsec:features}

To understand \alice in more detail, we first summarize its
basic features. \alice provides:

\begin{itemize}

\item polymorphic type constructors {\pisastyle{astream}} and
{\pisastyle{tastream}} for timed and untimed streams over
arbitrary HOL types,

\item sophisticated definition principles for streams and functions
over streams, such as pattern-matching,
recursion, and state-based definition techniques,

\item incorporated domain theory (concepts of approximation and
recursion),

\item various proof principles for streams,

\item incorporated automata constructs for state-based modeling,
also supporting underspecification or non-determinism,

\item extensive theories for handling timed streams, functions and properties,

\item a powerful simplifier (while developing \alice, a proper set of
simplification rules has been defined carefully in such a way as to be used 
by \alice automatically), and

\item an extensive library of functions on streams and theorems, as
well as commonly needed types
(just like in any other programming language, a good infrastructure
makes a language user friendly).

\end{itemize}

The following sections provide brief insights in the above listed features.
For a deeper understanding we refer to \cite{GR06}.

\delsecspace
\subsection{Specifying Streams}\label{subsec:streams}

\alice provides a basic type constructor called {\pisastyle{astream}} for
specifying untimed streams. For any Isabelle type {\pisastyle{t}},
the type {\pisastyle{t astream}} is member of the HOLCF type class
{\pisastyle{pcpo}} as described in Section \ref{subsec:holcf}.
The following exhaustion rule describes the basic structure of
untimed streams as well as the fundamental operators for their construction:

\begin{isabellebody}%
\sisanewline
{\isasymAnd}s{\isachardot}\ s\ {\isacharequal}\ {\isasymepsilon}\ {\isasymor}\ {\isacharparenleft}{\isasymexists}h\ rs{\isachardot}\ s\ {\isacharequal}\ {\isacharless}h{\isachargreater}{\isasymfrown}rs{\isacharparenright}
\sisanewline
\end{isabellebody}%

\noindent
A stream {\pisastyle{s}} is either empty, symbolized by {\isasymepsilon},
or there is a first message {\pisastyle{h}} and a remaining stream
{\pisastyle{rs}} so that pre-pending {\pisastyle{h}} to {\pisastyle{rs}}
yields the stream {\pisastyle{s}}. The operator {\pisastyle{<.>}} builds
single element streams and {\pisastyle{.{\isasymfrown}.}} defines
the concatenation on streams. It is associative and continuous in its
second argument and has the empty stream ({\isasymepsilon}) as a 
neutral element.
If the first argument of concatenation is
infinite, the second is irrelevant and the first is also the
result of the concatenation. This effectively means that the messages of
the second stream then never appear in the observation at all.

According to the above rules,
\alice also offers selection functions, named {\pisastyle{aft}} for
the head and {\pisastyle{art}} for the rest of a stream, respectively.
Function {\pisastyle{atake}} allows us to select the
first {\pisastyle{n}} symbols from a stream.
Function {\pisastyle{adrop}}
acts as a counterpart of {\pisastyle{atake}} as it drops the first
{\pisastyle{n}} messages from the beginning of a stream {\pisastyle{s}}.
The operator {\pisastyle{anth}} yields for a number
{\pisastyle{n}} and a stream {\pisastyle{s}}, the {\pisastyle{n}}-th message.
Beyond that, \alice provides many other auxiliary functions,
e.g. {\pisastyle{\isacharhash}} for the
length of a stream, yielding {\isasyminfinity} for infinite streams,
{\pisastyle{aflatten}} for the flattening of streams of streams,
{\pisastyle{aipower}} for the infinite repetition, 
{\pisastyle{afilter}} for message filtering.
In Section \ref{subsec:recfunctions} we give a tabular review of 
operators that are available in the current version of \alice.

Since streams are HOLCF datatypes, they carry a partial order 
(see also Section \ref{subsec:holcf}), which is described by
the following lemma

\begin{isabellebody}%
\sisanewline
s{\isadigit{1}}\ {\isasymsqsubseteq}\ s{\isadigit{2}}\ {\isasymLongrightarrow}\ {\isasymexists}t{\isachardot}\ s{\isadigit{1}}{\isasymfrown}t\ {\isacharequal}\ s{\isadigit{2}}
\sisanewline
\end{isabellebody}%

\noindent
The above rule characterizes the prefix ordering on streams.
It is induced by a flat order on the messages,
disregarding any internal structure of the messages themselves.
Based on these operators, a larger number of lemmas is
provided to deal with stream specifications, like case analysis,
unfolding rules, composition rules, associativity, injectivity, and
idempotency. Some foundational lemmas are given in Tab. \ref{table:streams}.

\makespace
\begin{table}
{\small
\caption{Some foundational lemmas on stream concatenation}
\label{table:streams}
\begin{center}
\begin{tabular}{|l|}
\hline
\spisastyle{
{\isasymepsilon}\hspace{0.05cm}{\isasymfrown}s = s{\isasymfrown}{\isasymepsilon} = s}\\
\spisastyle{
(s\isasymfrown t)\isasymfrown u = s\isasymfrown (t\isasymfrown u)}\qquad\\
\spisastyle{
\isacharhash{\isasymepsilon} = 0}\\
\spisastyle{
\isacharhash <m> = 1}\\
\spisastyle{
\isacharhash(s\isasymfrown t) = \isacharhash s + \isacharhash t}\\
\spisastyle{
\isacharhash s = \isasyminfinity\ \isasymLongrightarrow\ s\isasymfrown t = s}\\
\hline
\end{tabular}\end{center}}
\end{table}


\delsecspace
\subsection{Timed Streams}\label{subsec:tstreams}

Built on the untimed case,
\alice provides another type constructor called {\pisastyle{tastream}} for
specifying timed streams. Structurally, both are rather similar.
Again, for any Isabelle type {\pisastyle{t}},
the type {\pisastyle{t tastream}} is a member of {\pisastyle{pcpo}}.
The following exhaustion rule describes the basic structure of
timed streams. It shows that timed streams may still be empty,
contain a message or a tick as their first element:

\begin{isabellebody}%
\sisanewline
{\isasymAnd}ts{\isachardot}\ ts\ {\isacharequal}\ {\isasymepsilon}\ {\isasymor}\ {\isacharparenleft}{\isasymexists}z{\isachardot}\ ts\ {\isacharequal}\ {\isacharless}{\isasymsurd}{\isachargreater}{\isasymfrown}z{\isacharparenright}\ {\isasymor}\ {\isacharparenleft}{\isasymexists}m\ z{\isachardot}\ ts\ {\isacharequal}\ {\isacharless}Msg\ m{\isachargreater}{\isasymfrown}z{\isacharparenright}
\sisanewline
\end{isabellebody}%

\noindent
In addition to ordinary messages, we use a special message {\isasymsurd},
called the tick, to model time progress.
Each {\isasymsurd} stands for the end of a time frame.
To differentiate between the tick and ordinary messages, we use the
constructor {\pisastyle{Msg}} as shown above. This operator
is introduced by type constructor {\pisastyle{addTick}}
that extends any type with the tick.

Please note that any timed stream of type {\pisastyle{t tastream}}
is also an ordinary stream of type {\pisastyle{(t addTick) astream}}.
Therefore, all machinery for {\pisastyle{astream}} types is available.

In addition, \alice provides a timed take
function. {\pisastyle{ttake n\isasymcdot}s}
yields at most {\pisastyle{n}} time frames from the beginning of
a timed stream {\pisastyle{s}}.

To allow inductive definitions, {\pisastyle{tastream}} streams may
be empty. However, for specifications we restrict ourselves to
observations over infinite time, which means that we will
only use the subset of timed streams with infinitely many ticks.
Therefore, additional machinery is necessary to deal with those.
For example, the predicate {\pisastyle{timeComplete}} is provided
to check whether a stream contains infinitely many time frames.

For an integration of both stream classes,
operator {\pisastyle{timeAbs}} maps a timed stream into an untimed one,
just keeping the messages, but removing any time information.

\delsecspace
\subsection{Stream Based Proof Principles}\label{subsec:proofprinciples}

Having the necessary types and type classes as well as auxiliary functions
and lemmas at hand, we can introduce proof principles for streams now. At first,
we handle the untimed case, as the timed case can be built on that.

\delsecspace
\delsecspace
\subsubsection{Proof Principles for Untimed Streams.}

A rather fundamental proof principle for untimed streams is the so called
take-lemma for streams that gives us an inductive technique
for proving equality

\begin{isabellebody}%
\sisanewline
{\isacharparenleft}{\isasymforall}n{\isachardot}\ atake\ n{\isasymcdot}x\ {\isacharequal}\ atake\ n{\isasymcdot}y{\isacharparenright}\ {\isasymLongrightarrow}\ x\ {\isacharequal}\ y
\sisanewline
\end{isabellebody}%

\noindent
Two streams are equal if all
finite prefixes of the same length of the streams are equal.
More sophisticated proof principles, like pointwise comparison of two streams
using the operator {\pisastyle{anth}}
or the below given induction principles are built on the take-lemma.
The following is an induction principle for proving a property
{\pisastyle{P}} over finite (indicated by the constructor {\pisastyle{Fin}}) 
streams

\begin{isabellebody}%
\sisanewline
{\isasymlbrakk}{\isacharhash}x\ {\isacharequal}\ Fin\ n{\isacharsemicolon}\ P\ {\isasymepsilon}{\isacharsemicolon}\ {\isasymAnd}a\ s{\isachardot}\ P\ s\ {\isasymLongrightarrow}\ P\ {\isacharparenleft}{\isacharless}a{\isachargreater}{\isasymfrown}s{\isacharparenright}{\isasymrbrakk}\ {\isasymLongrightarrow}\ P\ x
\sisanewline
\end{isabellebody}%

\noindent
As said, when necessary, we base our proof principles directly on HOLCF
but try to avoid their extensive exposure. Here is a principle that
uses admissibility from HOLCF ({\pisastyle{adm}}) for predicates to
span validity to infinite streams (see \cite{Reg94a})

\begin{isabellebody}%
\sisanewline
{\isasymlbrakk}adm\ P{\isacharsemicolon}\ P\ {\isasymepsilon}{\isacharsemicolon}\ {\isasymAnd}a\ s{\isachardot}\ P\ s\ {\isasymLongrightarrow}\ P\ {\isacharparenleft}{\isacharless}a{\isachargreater}{\isasymfrown}s{\isacharparenright}{\isasymrbrakk}\ {\isasymLongrightarrow}\ P\ x
\sisanewline
\end{isabellebody}%

\noindent
The above induction principles have also been extended to
the general use of concatenation, where not only
single element streams, but arbitrary streams can be concatenated.

The concept of approximation (provided by HOLCF) and induction on
natural numbers can also be used to prove properties involving continuous
functions over streams as discussed in Section \ref{subsec:recfunctions}.

\delsecspace
\delsecspace
\subsubsection{Proof Principles for Timed Streams.}

Since timed streams can also be seen as normal untimed streams, the above
given proof principles can also be used to prove properties of timed streams.

Please note that we have taken a {\it message driven} approach to
inductively define timed streams. Messages are added individually
to extend a stream. This also leads to event driven specification techniques.
In the contrary, it would have been possible to model timed streams inductively
as a stream {\pisastyle{(t list) astream}}, where each list
denotes the finite list of messages of type {\pisastyle{t}} 
occurring in one time frame.
This definition would lead to time-driven specification principles.
It is up to further investigation to understand and integrate both
approaches. As a first step in this direction, \alice provides a
timed-take-lemma for timed streams arguing that streams are equal
if they are within first n time frames for each n, as given in the following.

\begin{isabellebody}%
\sisanewline
{\isacharparenleft}{\isasymforall}n{\isachardot}\ ttake\ n{\isasymcdot}x\ {\isacharequal}\ ttake\ n{\isasymcdot}y{\isacharparenright}\ {\isasymLongrightarrow}\ x\ {\isacharequal}\ y
\sisanewline
\end{isabellebody}%

\noindent
Analogously, the following proof principle is based on time frame comparison

\begin{isabellebody}%
\sisanewline
{\isacharparenleft}{\isasymforall}n{\isachardot}\ tframe\ n{\isasymcdot}x\ {\isacharequal}\ tframe\ n{\isasymcdot}y{\isacharparenright}\ {\isasymLongrightarrow}\ x\ {\isacharequal}\ y
\sisanewline
\end{isabellebody}%

\noindent
\alice provides more sophisticated proof principles for timed streams, but
also for special cases
of timed streams, such as time-synchronous streams, containing exactly one
message per time unit, and the already mentioned
time complete streams, containing infinitely many
time frames.

\delsecspace
\subsection{Recursive Functions on Streams}\label{subsec:recfunctions}

Specifying streams allows us to define observations on communication channels.
However, \alice focusses on specification of components
communicating over those channels. The behavior of
a component is generally modeled as function over streams
and is often defined recursively or even state-based.

A recursively defined function {\pisastyle{f}}
processes a prefix of its input stream {\pisastyle{s}} by
producing a piece of the output stream and continues to process the
remaining part of {\pisastyle{s}} recursively.
All functions defined in this specification style are per
construction correct behaviors for distributed components.
This makes such a specification style rather helpful.
Functions of this kind are defined in their simplest form as illustrated
in the following (using the function {\pisastyle{out}} to process 
the message {\pisastyle{x}} appropriately)

\begin{isabellebody}
\sisanewline
f (<x>{\isasymfrown}s) = (out x){\isasymfrown}(f s) 
\sisanewline
\end{isabellebody}

\noindent
By construction, these functions are monotonic and continuous 
(lub-preserving, see below) wrt. their inputs, 
which allows us to define a number of proof principles on functions.

\makespace
\begin{table}
{\small
\caption{Basic operators in \alice}
\label{table:operators1}
\begin{center}
\makespace
\begin{tabular}{|ll|}
\hline
Operator & Signature\\
\hline
\spisastyle{<.>} & \spisastyle{{\isacharprime}a\ 
{\isasymRightarrow}\ {\isacharprime}a\ astream}\\
\spisastyle{aft} & \spisastyle{{\isacharprime}a\ astream\ {\isasymRightarrow}\ 
{\isacharprime}a}\\
\spisastyle{art} & \spisastyle{{\isacharprime}a\ astream\ {\isasymrightarrow}\ 
{\isacharprime}a\ astream}\\
\spisastyle{atake} & \spisastyle{nat \isasymRightarrow\ 
{\isacharprime}a\ astream\ {\isasymrightarrow}\ 
{\isacharprime}a\ astream}\\
\spisastyle{adrop} & \spisastyle{nat \isasymRightarrow\ 
{\isacharprime}a\ astream\ {\isasymrightarrow}\ 
{\isacharprime}a\ astream} \\
\spisastyle{anth} & \spisastyle{nat \isasymRightarrow\ 
{\isacharprime}a\ astream\ {\isasymRightarrow}\ 
{\isacharprime}a} \\
\spisastyle{\isacharhash .}& \spisastyle{{\isacharprime}a\ astream\ 
{\isasymrightarrow}\ inat} \\
\spisastyle{.\isasymfrown .}& \spisastyle{{\isacharprime}a\ astream 
\isasymRightarrow\ {\isacharprime}a\ astream\ {\isasymrightarrow}\ 
{\isacharprime}a\ astream} \\
\spisastyle{aipower} & \spisastyle{{\isacharprime}a\ astream\ 
{\isasymRightarrow}\ {\isacharprime}a\ astream}\\
\spisastyle{apro1} & \spisastyle{{\isacharparenleft}{\isacharprime}a\ {\isacharasterisk}\ 
{\isacharprime}b{\isacharparenright}\ astream\ {\isasymrightarrow}\ 
{\isacharprime}a\ astream}\\
\spisastyle{apro2} & \spisastyle{{\isacharparenleft}{\isacharprime}a\ {\isacharasterisk}\ 
{\isacharprime}b{\isacharparenright}\ astream\ {\isasymrightarrow}\ 
{\isacharprime}b\ astream}\\
\spisastyle{amap} & {\spisastyle{{\isacharparenleft}{\isacharprime}a\ {\isasymRightarrow}\ 
{\isacharprime}b{\isacharparenright}\ {\isasymRightarrow}\ 
{\isacharprime}a\ astream\ {\isasymrightarrow}\ 
{\isacharprime}b\ astream}} \\
\spisastyle{azip} & {\spisastyle{{\isacharprime}a\ astream\ {\isasymrightarrow}\ 
{\isacharprime}b\ astream\ {\isasymrightarrow}\ 
{\isacharparenleft}{\isacharprime}a\ {\isacharasterisk}\ 
{\isacharprime}b{\isacharparenright}\ 
astream}} \\
\spisastyle{afilter} & {\spisastyle{{\isacharprime}a\ set\ {\isasymRightarrow}\ 
{\isacharprime}a\ astream\ {\isasymrightarrow}\ {\isacharprime}a\ astream}}\\
\spisastyle{atakew} & {\spisastyle{{\isacharparenleft}{\isacharprime}a\ 
{\isasymRightarrow}\ bool{\isacharparenright}\ 
{\isasymRightarrow}\ {\isacharprime}a\ astream\ {\isasymrightarrow}\ 
{\isacharprime}a\ astream}} \\
\spisastyle{adropw} & {\spisastyle{{\isacharparenleft}{\isacharprime}a\ 
{\isasymRightarrow}\ bool{\isacharparenright}\ 
{\isasymRightarrow}\ {\isacharprime}a\ astream\ {\isasymrightarrow}\ 
{\isacharprime}a\ astream}} \\
\spisastyle{aremstutter} & \spisastyle{{\isacharprime}a\ astream\ 
{\isasymrightarrow}\ {\isacharprime}a\ astream} \\
\spisastyle{aflatten} & \spisastyle{{\isacharprime}a\ astream\ astream\ 
{\isasymrightarrow}\ {\isacharprime}a\ astream} \\
\spisastyle{ascanl} & \spisastyle{nat \isasymRightarrow\ 
{\isacharparenleft}{\isacharprime}a\ {\isasymRightarrow}\ {\isacharprime}b\ 
{\isasymRightarrow}\ {\isacharprime}a{\isacharparenright}\ {\isasymRightarrow}\ 
{\isacharprime}a\ {\isasymRightarrow}\ {\isacharprime}b\ astream\ 
{\isasymrightarrow}\ {\isacharprime}a\ astream} \\
\spisastyle{aiterate} & \spisastyle{\isacharparenleft\isacharprime a\ {\isasymRightarrow}\ 
{\isacharprime}a{\isacharparenright}\ {\isasymRightarrow}\ 
{\isacharprime}a\ {\isasymRightarrow}\ 
{\isacharprime}a\ astream}\\
\hline
\end{tabular}\end{center}}
\end{table}

\makespace
\delsecspace

A number of predefined auxiliary operators assist in specifying components.
Due to expressiveness, we also allow to use
operators that are not monotonic or continuous in some arguments,
such as {\isasymfrown} in its first argument or {\pisastyle{aipower}}. 
In \alice, it is also possible to define more such functions using 
pattern-matching and recursion. The above notions can also be found in 
standard literature on semantics like \cite{Win93}.
In the following we concentrate on continuous functions.

\delsecspace
\delsecspace
\subsubsection{Continuous Functions - The Approximation Principle.}

As briefly discussed, continuous functions capture the notion of
computability in interactive
systems and therefore play a prominent role in stream-processing
specification techniques.
The behavior of a continuous function for an infinite input can be predicted
by the behavior for the finite parts of the input.
Thus, its behavior can be approximated.
As it has been shown amongst others in \cite{Win93},
composition of continuous functions results in continuous functions.
Therefore, based on a number of basic functions and equipped
with appropriate definition techniques, it becomes easy to
specify further functions. \alice provides amongst others

\begin{itemize}

\item pattern-matching and recursion (like in functional languages),

\item state-based definitions (using \io-automata \cite{Rum96}, see
Section \ref{subsec:statebased}),

\item fixpoint recursion (using HOLCF), and

\item continuous function-chain construction (using HOL's
{\pisastyle{primrec}} and approximation, see \cite{GR06})

\end{itemize}

\noindent
Currently, we do have at least the operators on streams
depicted in Tab. \ref{table:operators1} and Tab. \ref{table:operators2} 
available. 
For the sake of brevity, we do not explain those
further, but refer to \cite{GR06} as well as Section \ref{subsec:streams}
and \ref{subsec:tstreams}
and furthermore assume that readers will recognize the
functionality through name and signature.

\makespace
\begin{table}
{\small
\caption{Basic operators for timed specifications}
\label{table:operators2}
\begin{center}
\makespace
\begin{tabular}{|ll|}
\hline
Operator & Signature\\
\hline
\spisastyle{timeComplete} & \spisastyle{{\isacharprime}a\ 
tastream {\isasymRightarrow}\ bool}\\
\spisastyle{timeSync} & \spisastyle{{\isacharprime}a\ 
tastream {\isasymRightarrow}\ bool}\\
\spisastyle{injectTicks} & \spisastyle{nat astream {\isasymrightarrow}\ 
{\isacharprime}a\ astream {\isasymrightarrow}\ {\isacharprime}a tastream}\\
\spisastyle{timeAbs} & \spisastyle{{\isacharprime}a\ 
tastream {\isasymrightarrow}\ {\isacharprime}a\ astream}\\
\spisastyle{ttake} & \spisastyle{nat \isasymRightarrow\ 
{\isacharprime}a\ tastream\ {\isasymrightarrow}\ 
{\isacharprime}a\ tastream}\\
\spisastyle{tframe} & \spisastyle{nat \isasymRightarrow\ 
{\isacharprime}a\ tastream\ {\isasymrightarrow}\ 
{\isacharprime}a\ astream}\\
\spisastyle{stretchTimeFrame} & \spisastyle{{nat \isasymRightarrow\
\isacharprime}a\ tastream {\isasymrightarrow}\ {\isacharprime}a\ tastream}\\
\spisastyle{getTime} & \spisastyle{{nat \isasymRightarrow\ \isacharprime}a\ 
tastream {\isasymRightarrow}\ nat}\\
\hline
\end{tabular}\end{center}}
\end{table}

\makedoublespace

\delsecspace
\subsection{State-Based Definition Techniques}\label{subsec:statebased}

There is quite a number of variants of state machines available
that allow for a state-based description.
We use \io-automata that do have transitions with
one occurring message (event) as input and a sequence of messages
(events) as output (hence \io).
They have been defined in \cite{Rum96}
together with a formal semantics based on streams and
a number of refinement techniques.
In contrast to I/O automata \cite{LT89}, they couple
incoming event and reaction and need no intermediate states.

As they are perfectly suited for a state-based description of
component behavior, we provide assistance for the definition of
an \io-automaton {\pisastyle{A}}
in \alice by modeling the abstract syntax as a 5-tuple in form of

\begin{isabellebody}
\sisanewline
A = (stateSet A, inCharSet A, outCharSet A, delta A, initSet A)
\sisanewline
\end{isabellebody}

\noindent
Automata of this structure can be defined using the type constructor
{\pisastyle{ioa}}. \io-automata
consist of types for its states, input and
output messages. {\pisastyle{delta}}
denotes the transition relation of an automaton. It consists of tuples
of source state, input message, destination state and a sequence
of output messages.
The 5th element {\pisastyle{initSet}} describes start states and possible
initial output (that is not a reaction to any incoming message).

As an illustration, we define\footnote{Due to lack of space,
we skip HOL's keyword {\spisastyle{constdefs}} in front of a definition
but symbolize it by indentation.
We also do not introduce the necessary type declarations,
which is actually straightforward for the specifications used here.} 
an \io-automaton representing a
component dealing with auctions in the American style, where
bidders spontaneously and repeatedly spend money and after a certain 
(previously unknown) timeout the last spender gets the auctioned
artifact. The auction component is initialized with an arbitrary 
but a non-zero timeout. 
It counts down using the ticks and stores the last 
bidder as he will be the winner.

\begin{isabellebody}%
\sisanewline
\ \ amiauction\ {\isacharcolon}{\isacharcolon}\ {\isachardoublequoteopen}{\isacharparenleft}{\isacharparenleft}nat\ {\isacharasterisk}\ Bid\ {\isacharasterisk}\ IAP{\isacharparenright}{\isacharcomma}\ Bid\ addTick{\isacharcomma}\ BidUclosed\ addTick{\isacharparenright}\ ioa{\isachardoublequoteclose}\isanewline
\ \ amiauction{\isacharunderscore}def{\isacharcolon}\ \isanewline
\ \ \ \ {\isachardoublequoteopen}amiauction\ {\isasymequiv}\ \isanewline
\ \ \ \ \ \ {\isacharparenleft}UNIV{\isacharcomma}\ UNIV{\isacharcomma}\ UNIV{\isacharcomma}\ \isanewline
\ \ \ \ \ \ \ {\isacharbraceleft}t{\isachardot}{\isasymexists}k\ b\ m\ x{\isachardot}\ \isanewline
\sisanewline
\ \ \ \ \ \ \ \ \ \ \ {\isacharparenleft}{\isacharasterisk}\ handle\ time\ and\ accept\ the\ last\ bid\ \isanewline
\ \ \ \ \ \ \ \ \ \ \ \ \ \ \ \ as\ soon\ as\ the\ time\ limit\ is\ reached\ {\isacharasterisk}{\isacharparenright}\ \isanewline
\ \ \ \ \ \ \ \ \ \ \ t\ {\isacharequal}\ {\isacharparenleft}{\isacharparenleft}k{\isacharplus}{\isadigit{1}}{\isacharcomma}b{\isacharcomma}x{\isacharparenright}{\isacharcomma}\ {\isasymsurd}{\isacharcomma}\ {\isacharparenleft}k{\isacharcomma}b{\isacharcomma}x{\isacharparenright}{\isacharcomma}\ {\isacharless}{\isasymsurd}{\isachargreater}{\isacharparenright}\ {\isasymand}\ k\ {\isachargreater}\ {\isadigit{0}}\ {\isasymor}\isanewline
\ \ \ \ \ \ \ \ \ \ \ t\ {\isacharequal}\ {\isacharparenleft}{\isacharparenleft}{\isadigit{0}}{\isacharcomma}b{\isacharcomma}x{\isacharparenright}{\isacharcomma}\ {\isasymsurd}{\isacharcomma}\ {\isacharparenleft}{\isadigit{0}}{\isacharcomma}b{\isacharcomma}x{\isacharparenright}{\isacharcomma}\ {\isacharless}{\isasymsurd}{\isachargreater}{\isasymfrown}{\isacharless}Msg\ closed{\isachargreater}{\isacharparenright}\ {\isasymor}\isanewline
\ \ \ \ \ \ \ \ \ \ \ t\ {\isacharequal}\ {\isacharparenleft}{\isacharparenleft}{\isadigit{1}}{\isacharcomma}b{\isacharcomma}I{\isacharparenright}{\isacharcomma}\ {\isasymsurd}{\isacharcomma}\ {\isacharparenleft}{\isadigit{0}}{\isacharcomma}b{\isacharcomma}I{\isacharparenright}{\isacharcomma}\ {\isacharless}{\isasymsurd}{\isachargreater}{\isasymfrown}{\isacharless}Msg\ closed{\isachargreater}{\isacharparenright}\ {\isasymor}\isanewline
\ \ \ \ \ \ \ \ \ \ \ t\ {\isacharequal}\ {\isacharparenleft}{\isacharparenleft}{\isadigit{1}}{\isacharcomma}b{\isacharcomma}A{\isacharparenright}{\isacharcomma}\ {\isasymsurd}{\isacharcomma}\ {\isacharparenleft}{\isadigit{0}}{\isacharcomma}b{\isacharcomma}A{\isacharparenright}{\isacharcomma}\ {\isacharless}{\isasymsurd}{\isachargreater}{\isasymfrown}{\isacharless}Msg\ {\isacharparenleft}accept\ b{\isacharparenright}{\isachargreater}{\isasymfrown}{\isacharless}Msg\ closed{\isachargreater}{\isacharparenright}\ {\isasymor}\isanewline
\ \ \ \ \ \ \ \ \ \ \ \sisanewline
\ \ \ \ \ \ \ \ \ \ \ {\isacharparenleft}{\isacharasterisk}\ store\ the\ new\ bid\ m\ if\ necessary\ {\isacharasterisk}{\isacharparenright}\isanewline
\ \ \ \ \ \ \ \ \ \ \ t\ {\isacharequal}\ {\isacharparenleft}{\isacharparenleft}k{\isacharplus}{\isadigit{1}}{\isacharcomma}b{\isacharcomma}x{\isacharparenright}{\isacharcomma}\ Msg\ m{\isacharcomma}\ {\isacharparenleft}k{\isacharplus}{\isadigit{1}}{\isacharcomma}m{\isacharcomma}A{\isacharparenright}{\isacharcomma}\ {\isasymepsilon}{\isacharparenright}\ {\isasymor}\ t\ {\isacharequal}\ {\isacharparenleft}{\isacharparenleft}{\isadigit{0}}{\isacharcomma}b{\isacharcomma}x{\isacharparenright}{\isacharcomma}\ Msg\ m{\isacharcomma}\ {\isacharparenleft}{\isadigit{0}}{\isacharcomma}b{\isacharcomma}x{\isacharparenright}{\isacharcomma}\ {\isasymepsilon}{\isacharparenright}{\isacharbraceright}{\isacharcomma}\isanewline
\sisanewline
\ \ \ \ \ \ \ {\isacharbraceleft}{\isacharparenleft}{\isacharparenleft}{\isasymsome}s{\isachardot}\ fst\ s\ {\isachargreater}\ {\isadigit{0}}\ {\isasymand}\ snd\ {\isacharparenleft}snd\ s{\isacharparenright}\ {\isacharequal}\ I{\isacharparenright}{\isacharcomma}\ {\isasymepsilon}{\isacharparenright}{\isacharbraceright}{\isacharparenright}{\isachardoublequoteclose}
\sisanewline
\end{isabellebody}%

\noindent
The above automaton is well-defined, deterministic and complete.
By applying the operator {\pisastyle{ioafp}}, we
map this automaton into a function that is continuous by construction.
The recursive definition of a stream-processing function is now
embedded in the {\pisastyle{ioafp}} operator, leaving a non-recursive
but explicit definition of the actual behavior in an event based style.

In fact, a number of proof principles are established on these state machines
that do not need inductive proof anymore, but just need to compare
transitions and states. More precisely, the behaviors can then be compared
by establishing a (bi-)simulation relation between the automata.

A non-deterministic \io-automaton is defined in an analogous form 
and not mapped to a single but a set of stream-processing functions.
This is especially suitable to deal with underspecification.

As said, \alice is still in development. Although we have
initial results on this kind of specification style, we will
further elaborate \alice to comfortably deal with
\io-automata of this kind in the future.

\delsecspace
\section{Alternating Bit Protocol - An Example}\label{sec:abp}

Based on the theory introduced so far, we show the usefulness
of \alice by developing a small, yet not trivial
and well known example.

The Alternating Bit Protocol (ABP) is a raw transmission protocol
for data over an unreliable medium. Goal of the ABP is to transmit
data over a medium that looses some messages, but does not create, modify,
rearrange or replicate them. The key idea is that the sender adds an 
identifier to each message that is being sent back as acknowledgement by the
receiver. If the acknowledgement does not arrive, the sender sends
the same message again. When only one single message is in
transmission, the identifier can boil down to a single bit with
alternating value -- hence the name of the protocol.

\begin{figure}
\begin{center}
\scalebox{0.5}{\includegraphics{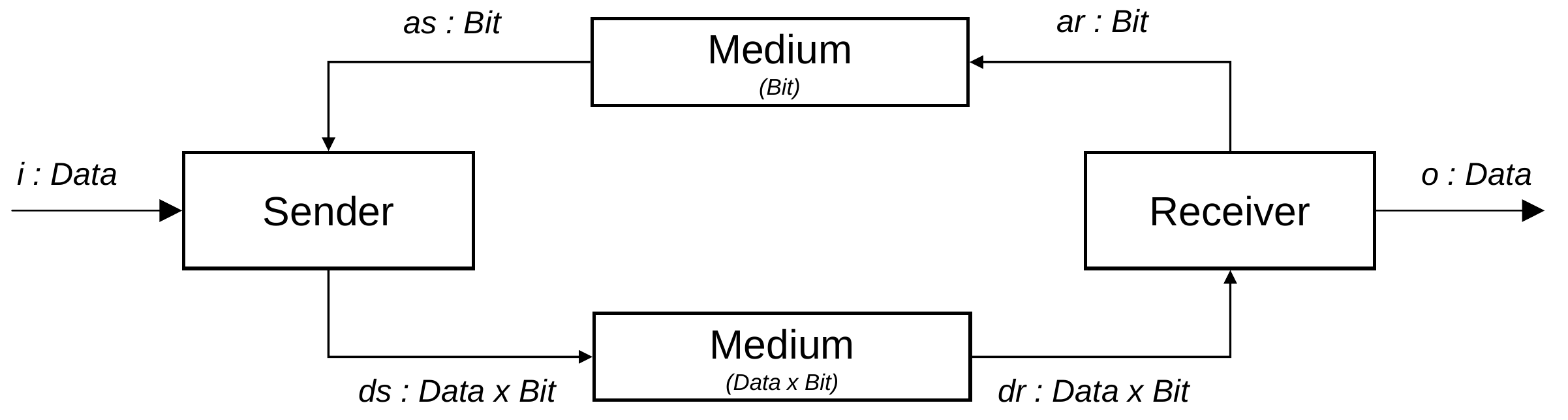}}
\caption{The architecture of the Alternating Bit Protocol (ABP)}
\label{fig:abp}
\end{center}
\end{figure}

The ABP specification involves a number of typical issues, such as
underspecification, unbounded non-determinism and fairness.
Fig. \ref{fig:abp} illustrates the overall structure of the ABP.
A detailed explanation of a similar specification can be found in \cite{BS01}.

\makespace
\makedoublespace

\delsecspace
\subsection{The ABP Medium}

Please note that the medium is modeled after the existing, real world,
while sender and receiver need to be specified and later implemented in
such a way that they can safely deal with the given medium.
So, we first specify the behavior of the medium as described above.

\begin{isabellebody}%
\sisanewline
\ \ Med\ {\isacharcolon}{\isacharcolon}\ {\isachardoublequoteopen}{\isacharprime}t\ astream\ {\isasymRightarrow}\ {\isacharprime}t\ astream\ {\isasymRightarrow}\ bool{\isachardoublequoteclose}\isanewline
\ \ Med{\isacharunderscore}def{\isacharcolon}\ \ \isanewline
\ \ \ \ {\isachardoublequoteopen}Med\ x\ y\ {\isasymequiv}\ \isanewline
\ \ \ \ \ \ \ {\isasymexists}p{\isachardot}\ {\isacharhash}{\isacharparenleft}afilter\ {\isacharbraceleft}True{\isacharbraceright}{\isasymcdot}p{\isacharparenright}\ {\isacharequal}\ {\isasyminfinity}\ {\isasymand}\ \isanewline
\ \ \ \ \ \ \ \ \ \ \ y\ {\isacharequal}\ apro{\isadigit{1}}{\isasymcdot}{\isacharparenleft}afilter\ {\isacharbraceleft}a{\isachardot}\ {\isasymexists}b{\isachardot}\ a\ {\isacharequal}\ {\isacharparenleft}b{\isacharcomma}\ True{\isacharparenright}{\isacharbraceright}{\isasymcdot}{\isacharparenleft}azip{\isasymcdot}x{\isasymcdot}p{\isacharparenright}{\isacharparenright}{\isachardoublequoteclose}\ \sisanewline
\end{isabellebody}%

\noindent
Through the use of an internal oracle stream {\pisastyle{p}},
we can describe that a medium does eventually transmit a message if we retry
long enough.
The fairness, as described below, is deduced from the
above specification as follows.

\begin{isabellebody}%
\sisanewline
{\isasymlbrakk}{\isacharhash}x\ {\isacharequal}\ {\isasyminfinity}{\isacharsemicolon}\ Med\ x\ y{\isasymrbrakk}\ {\isasymLongrightarrow}\ {\isacharhash}y\ {\isacharequal}\ {\isasyminfinity}\sisanewline
\end{isabellebody}%

\noindent
The lemma is proven easily using the following auxiliary lemma, since
the lengths of the first and the second pointwise projection 
({\pisastyle{apro1}} and {\pisastyle{apro2}} respectively)
of a stream consisting of ordered pairs are equal.

\begin{isabellebody}%
\sisanewline
{\isasymforall}x{\isachardot}\ {\isacharhash}x\ {\isacharequal}\ {\isasyminfinity}\ {\isasymlongrightarrow}\ 
apro{\isadigit{2}}{\isasymcdot}{\isacharparenleft}afilter\ {\isacharbraceleft}a{\isachardot}\ {\isasymexists}b{\isachardot}\ a\ {\isacharequal}\ {\isacharparenleft}b{\isacharcomma}\ z{\isacharparenright}{\isacharbraceright}{\isasymcdot}{\isacharparenleft}azip{\isasymcdot}x{\isasymcdot}p{\isacharparenright}{\isacharparenright}\ {\isacharequal}\ afilter\ {\isacharbraceleft}z{\isacharbraceright}{\isasymcdot}p\sisanewline
\end{isabellebody}%

\noindent
The above auxiliary lemma is again proven by induction on the free stream
variable {\pisastyle{p}} using an appropriate
proof principle from Section \ref{subsec:proofprinciples}.

\delsecspace
\subsection{The Sender}

Now, relative to a given medium, we have to define a sender and a
receiver that establish the desired behavior: safe transmission of
messages.
The sender receives data from outside and transmits them together
with the alternating bit. We give a specification in a functional style:

\begin{isabellebody}%
\sisanewline
\ \ Snd\ {\isacharcolon}{\isacharcolon}\ {\isachardoublequoteopen}Data\ astream\ {\isasymRightarrow}\ Bit\ astream\ {\isasymRightarrow}\ {\isacharparenleft}Data\ {\isacharasterisk}\ Bit{\isacharparenright}\ astream\ {\isasymRightarrow}\ bool{\isachardoublequoteclose}\isanewline
\ \ Snd{\isacharunderscore}def{\isacharcolon}\ \isanewline
\ \ \ \ {\isachardoublequoteopen}Snd\ i\ as\ ds\ {\isasymequiv}\ \isanewline
\ \ \ \ \ \ \ let\ \isanewline
\ \ \ \ \ \ \ \ \ \ fas\ {\isacharequal}\ aremstutter{\isasymcdot}as{\isacharsemicolon}\ \isanewline
\ \ \ \ \ \ \ \ \ \ fb\ {\isacharequal}\ apro{\isadigit{2}}{\isasymcdot}{\isacharparenleft}aremstutter{\isasymcdot}ds{\isacharparenright}{\isacharsemicolon}\isanewline
\ \ \ \ \ \ \ \ \ \ fds\ {\isacharequal}\ apro{\isadigit{1}}{\isasymcdot}{\isacharparenleft}aremstutter{\isasymcdot}ds{\isacharparenright}\ \isanewline
\ \ \ \ \ \ \ in\ \isanewline
\ \ \ \ \ \ \ \ \ \ fds\ {\isasymsqsubseteq}\ i\ {\isasymand}\ \isanewline
\ \ \ \ \ \ \ \ \ \ fas\ {\isasymsqsubseteq}\ fb\ {\isasymand}\ \ \ \isanewline
\ \ \ \ \ \ \ \ \ \ aremstutter{\isasymcdot}fb\ {\isacharequal}\ fb\ {\isasymand}\ \ \isanewline
\ \ \ \ \ \ \ \ \ \ {\isacharhash}fds\ {\isacharequal}\ imin\ {\isacharhash}i\ {\isacharparenleft}iSuc\ {\isacharparenleft}{\isacharhash}fas{\isacharparenright}{\isacharparenright}\ {\isasymand}\isanewline
\ \ \ \ \ \ \ \ \ \ {\isacharparenleft}{\isacharhash}fas\ {\isacharless}\ {\isacharhash}i\ {\isasymlongrightarrow}\ {\isacharhash}ds\ {\isacharequal}\ {\isasyminfinity}{\isacharparenright}{\isachardoublequoteclose}\sisanewline
\end{isabellebody}%

\noindent
We explicitly define the channel observations for the sender 
in Fig. \ref{fig:abp}.
The conjuncts in the {\pisastyle{in}} part of the definition 
constrain the sender in the order of their appearance, using the
abbreviations from the {\pisastyle{let}} part, as follows
\begin{enumerate}

\item Abstracting from consecutive repetitions of a message via
{\pisastyle{aremstutter}}, we see that the sender is sending
the input messages in the order they arrive.

\item The sender also knows which acknowledgement
bit it is waiting for, nevertheless, it is underspecified
which acknowledgment bit is sent initially.

\item Each new element from the data input channel is assigned a bit different
from the bit previously assigned.

\item When an acknowledgment is received, the next data element will
eventually be transmitted, given that there are more data elements
to transmit.

\item If a data element is never acknowledged then the sender never stops
transmitting this data element. 
\end{enumerate}

\delsecspace
\subsection{The Receiver}

The receiver sends each acknowledgment bit back to the
sender via the acknowledgment medium and the received data messages to the
data output channel removing consecutive repetitions, respectively.

\begin{isabellebody}%
\sisanewline
\ \ Rcv\ {\isacharcolon}{\isacharcolon}\ {\isachardoublequoteopen}{\isacharparenleft}Data\ {\isacharasterisk}\ Bit{\isacharparenright}\ astream\ {\isasymRightarrow}\ Bit\ astream\ {\isasymRightarrow}\ Data\ astream\ {\isasymRightarrow}\ bool{\isachardoublequoteclose}\isanewline
\ \ Rcv{\isacharunderscore}def{\isacharcolon}\ 
{\isachardoublequoteopen}Rcv\ dr\ ar\ o\ {\isasymequiv}\ 
ar\ {\isacharequal}\ apro{\isadigit{2}}{\isasymcdot}dr\ {\isasymand}\ 
o\ {\isacharequal}\ apro{\isadigit{1}}{\isasymcdot}{\isacharparenleft}aremstutter{\isasymcdot}dr{\isacharparenright}{\isachardoublequoteclose}\sisanewline
\end{isabellebody}%

\delsecspace
\subsection{The Composed System}

The overall system is composed as defined by the architecture in Fig.
\ref{fig:abp}. This composition is straightforwardly to formulate in \alice:

\begin{isabellebody}%
\sisanewline
\ \ ABP\ {\isacharcolon}{\isacharcolon}\ {\isachardoublequoteopen}Data\ astream\ {\isasymRightarrow}\ Data\ astream\ {\isasymRightarrow}\ bool{\isachardoublequoteclose}\isanewline
\ \ ABP{\isacharunderscore}def{\isacharcolon}\ \isanewline
\ \ \ \ {\isachardoublequoteopen}ABP\ i\ o\ {\isasymequiv}\ {\isasymexists}as\ ds\ dr\ ar{\isachardot}\ 
Snd\ i\ as\ ds\ {\isasymand}\ Med\ ds\ dr\ {\isasymand}\ Rcv\ dr\ ar\ o\ {\isasymand}\ Med\ ar\ as{\isachardoublequoteclose}\ \sisanewline
\end{isabellebody}%


This formalization of the ABP uses a relational approach similar to the
specification in \cite{BS01}. However, formalizations as sets of functions
or in a state-based manner are possible as well. Using a more elaborate
version of \alice, we will be able to define a state-based version of sender
and receiver (similar to \cite{GGR06}), 
which is on the one hand more oriented towards implementation and
on the other hand might be more useful for inductive proof on the
behaviors.
Most important however, we will be able to prove that this relational
and the state-based specifications will coincide.

For this case study, we remain in the relational style and
specify the expected property of the overall system 
(without actually presenting the proof):

\begin{isabellebody}%
\sisanewline
ABP\ i\ o\
{\isasymLongrightarrow}\ o\ {\isacharequal}\ i
\sisanewline
\end{isabellebody}%

\noindent
Please note that, at this stage of the development of ABP,
there are neither realizability nor sophisticated timing constraints
considered in the above formalization.
Due to relational semantics,
additional refinement steps are then needed to reduce
the underspecification towards an implementation oriented or timing-aware 
style, since there are infinite streams fulfilling the specification 
that are not valid protocol histories. These, however, would not occur,
when using sets of stream-processing functions as they impose continuity
on the overall behavior.

\delsecspace
\section{Discussion}
\label{sec:lastsec}

In this paper we have introduced \alice, an advanced logic for formal
specification and verification of communication in distributed systems.
\alice is embedded in the higher order logic HOL, 
which itself is formalized using the Isabelle
generic theorem prover.

Our approach is based on using HOLCF to deal with
partiality, infinity, recursion, and continuity.
We provide techniques to use \alice directly from HOL, thus
preventing the user to actually deal with HOLCF specialities.

\alice is currently under development. So not all concepts
and theories presented here are already completely mature.
Further investigations will also deal with the question of expressiveness,
applicability and interoperability. 
Beyond the ABP, we already have some experience
with other formalizations that show that the overhead
of formalizing a specification in \alice as apposed to a mere paper
definition is not too bad. 
However, it also shows where to improve comfort.


\delsecspace
\bibliographystyle{alpha}
\bibliography{refs}

\end{document}